\documentclass[a4paper,11pt]{article}

\usepackage{jcappub} 
\usepackage[T1]{fontenc} 
\usepackage{graphics}
\usepackage{breakurl}

\title{\boldmath Constraints on Yukawa gravity parameters from observations of bright stars}

\author[a,1]{P. Jovanovi\'{c},\note{Corresponding author.}}
\author[b]{V. Borka Jovanovi\'{c},}
\author[b]{D. Borka}
\author[c]{and A. F. Zakharov}

\affiliation[a]{Astronomical Observatory, Volgina 7, P.O. Box 74, 11060 Belgrade, Serbia}
\affiliation[b]{Department of Theoretical Physics and Condensed Matter Physics (020), \\
Vin\v{c}a Institute of Nuclear Sciences - National Institute of the Republic of Serbia, \\
University of Belgrade, P.O. Box 522, 11001 Belgrade, Serbia}
\affiliation[c]{Bogoliubov Laboratory for Theoretical Physics, JINR, 141980 Dubna, Russia}

\emailAdd{pjovanovic@aob.rs}
\emailAdd{vborka@vinca.rs}
\emailAdd{dusborka@vinca.rs}
\emailAdd{alex.fed.zakharov@gmail.com}

\abstract{In this paper we investigate a Yukawa gravity modification of the Newtonian gravitational potential in a weak field approximation. For that purpose we derived the corresponding equations of motion and used them to perform two-body simulations of the stellar orbits. In 2020 the GRAVITY Collaboration detected the orbital precession of the S2 star around the supermassive black hole (SMBH) at the Galactic Center (GC) and showed that it is close to the general relativity (GR) prediction. Using this observational fact, we evaluated parameters of the Yukawa gravity (the range of Yukawa interaction $\Lambda$ and universal constant $\delta$) with the Schwarzschild precession of the S-stars assuming that the observed values as indicated by the GRAVITY Collaboration will have a small deviation from GR prediction \cite{abut20}. GR provides the most natural way to fit observational data for S-star  orbits, however, their precessions can be fitted by Yukawa gravity. Our main goal was to study the possible influence of the strength of Yukawa interaction, i.e. the universal constant $\delta$, on the precessions of S-star orbits. We analyze S-star orbits assuming different strength of Yukawa interaction $\delta$ and find that this parameter has strong influence on range of Yukawa interaction $\Lambda$. For that purpose we use parameterized post-Newtonian (PPN) equations of motion in order to calculate the simulated orbits of S-stars in GR and Yukawa gravity. Using MCMC simulations we obtain the best-fit values and uncertainties of Yukawa gravity parameters for S-stars. Also, we introduce a new criterion which can be used for classification of gravitational systems in this type of gravity, according to their scales. We demonstrated that performed analysis of the observed S-stars orbits around the GC in the frame of the Yukawa gravity represent a tool for constraining the Yukawa gravity parameters and probing the predictions of gravity theories.}

\begin{document}
\maketitle
\flushbottom

\section{Introduction}

General Relativity (GR) was introduced by A. Einstein in 1915 and consequent observations and experiments showed that GR is the best gravity model \cite{damo06,cho11,paol22}, since GR successfully passed all considered tests.

The Galactic Center (radio source Sgr A*) is the closest galactic center. Astronomers believe that supermassive black hole (SMBH) is located at the Galactic Center (GC) as it was suggested in \cite{lynd71,oort77,rees82} however, many alternative models have been proposed for GC, including for instance a dense cluster of stars \cite{reid09}, fermion ball \cite{muny02}, boson stars \cite{jetz92,torr00,vinc16}, neutrino balls \cite{viol93,paol01} and others \cite{muny98,tsik98,bili98,bili02}. Recently, it was proposed to substitute the SMBH at GC by dark matter concentration with a dense core and diluted halo \cite{ruff15} and it was claimed that such an approach provides a better fit in comparison of the conventional approach where a SMBH is the key component \cite{bece21}. However, a consequent analysis of the RAR-model (Ruffini, Arguelles, Rueda) showed that since in spite of the fact that trajectories of bright stars inside a ball with a constant density are elliptical their properties are different from observational ones \cite{zakh22,zakh22b} and therefore, to fit trajectories of bright stars almost all dark matter mass has to be in a ball with a radius which is smaller than pericenters of these stars \cite{argu22}. Therefore, many alternatives for SMBH were significantly constrained by consequent observations and SMBH is the most reliable and natural model for GC. The nearest SMBH is located in the center of our Galaxy; therefore, this object is very attractive for observations, and astronomers observe it in various spectral ranges, including gamma-ray, X-ray, infrared, optical and radio ranges.

About 50 years ago Bardeen \cite{bard73} presented a picture of a dark region (a shadow) for a thought observation which corresponds to a bright screen located behind a Kerr black hole and a distant observer is located in the equatorial plane. Later, Chandrasekhar \cite{chan83} reproduced a similar picture in his book. However, neither Bardeen nor Chandrasekhar did not consider shadow as a possible test of GR since a) shadow sizes are extremely small to be detected for known black holes and b) there are no bright screens precisely behind black hole in astronomy. In addition we would like to note that perhaps  Bardeen and Chandrasekhar did not suggest to use the apparent shape of a black hole as GR test, because the dark region (shadow) is too small to be detectable for all known estimates of black hole masses and distances toward them. In papers \cite{falc00,meli01} the authors showed results of simulations of  shadow formation for the GCin the framework of a numerical model, where the authors took into account electron scattering for a radiation in mm and cm bands. The authors concluded that  it is possible to observe  a dark region (shadow) around the black hole  in mm band, while it is not possible to see a shadow in cm band due to electron scattering. It was expected to create a global network acting in 1.3~mm wavelength, therefore the best angular resolution of this interferometer is around $25~\mu as$ (similar to the current resolution of Event Horizon Telescope (EHT) network, while the shadow diameter was estimated as small as $30~\mu as$ assuming that the black hole mass is $2.6 \times 10^6M_\odot$ as it was evaluated in \cite{ecka96,ghez98}, therefore, expectations for shadow observations with these facilities were not very optimistic, however, now we know that the black hole mass is more than $4 \times 10^6M_\odot$ and EHT Collaboration reconstructed the shadow at Sgr A* in 2022. Therefore, a problem of shadow reconstruction using EHT Collaboration observations is very hard but slightly simpler than it was expected in 2000.

At the beginning of this century, astrophysical problems were discussed for the Radioastron  space telescope under construction; it was supposed to conduct simultaneous observations with ground based telescopes. The angular resolution  of such an Earth--space interferometer could be of the order of 7 angular microseconds at the shortest wavelength of 1.3 cm. The mass of the  black hole in the GC is about $4 \times 10^6~M_\odot$, and the distance to it is of the order of 8 kpc; thereby, the angular size of the Schwarzschild radius is $10~\mu as$, which turns out to be comparable with the angular resolution of the interferometer. It would seem that we  could discuss manifestations of the general theory of relativity that could be detected. Assuming that the photons are only curved by the gravitational field of the black hole, but do not scatter near its vicinity, the size and shape of the shadow (that is, the dark area for the distant observer) was described in \cite{zakh05,zakh05b}. This means that for the observer standing in the equatorial plane, the value of is the same (it does not depend on the spin value), and the shadow is deformed in the direction parallel to the equatorial plane; the deformation depends on spin. Thus, the size of the shadow for the black hole in the GC is about $50~\mu as$ and, in principle, the structures with such a size could be observed with the Radioastron. However, scientists subsequently understood that, apparently, the image shape is smeared due to the scattering of photons on electrons for the centimeter range, and in order to detect the shadow, it is necessary to switch to the millimeter and submillimeter wavelength ranges, i.e., to the wavelengths that are planned to be used in the designed Millimetron space interferometer, as noted in \cite{zakh05,zakh05b}. An international global network of telescopes, including the millimeter wavelength range or the so-called the Event Horizon Telescope, operating as a giant interferometer the size of the Earth, observes the distribution of bright spots in the vicinity of the GC and determines the parameters of the black hole in the GC, and constrains alternative theories of gravity using this distribution. In April 2019, this team of scientists reported the first image of a supermassive black hole in the center of the M87 galaxy, which is located at a distance of about 17 Mpc from the Earth and the mass of the black hole in its center is (in April 2017, observations were made at a wavelength of 1.3 mm). The shadow size for this black hole is about $42~\mu as$ \cite{akiy19}, that is, the black hole in the center of the M87 galaxy is much more massive and it is located much farther than the GC, but their shadows are similar in size (this situation is somewhat similar that the angular sizes of the Sun and Moon visible from the Earth are very close in size) since it was shown recently by the EHT collaboration the shadow diameter for Sgr A* is $52~\mu as$ \cite{akiy22}.

As it was noted earlier to evaluate the parameters of the gravitational potential in the GC, photons may be used as test particles (in this case, the size and shape of the shadow in the vicinity of the GC are analyzed). Also, the bright stars that move in the vicinity of the black hole are used as test particles. Despite the fact that GR predictions are confirmed by the results of many observations and experiments (in particular, for the  gravitational field in the vicinity of a black hole in the GC in 2018, the GRAVITY collaboration confirmed GR predictions about the gravitational redshift of the spectrum of star S2 near the pericenter passage on May 18, 2018 and these conclusions are in good agreement with the observations presented by the Keck team \cite{abut20,abut18,abut19,do19}). Below we will use observations of bright star trajectories to constrain gravitational potential at Sgr A*.

As the carrier of the gravitational interaction, graviton is considered to be spin-2 (tensor) boson, electrically uncharged, as well as massless since, according to GR, it travels along null geodesics at the speed of light $c$ (like photon). However, according to some alternative theories, gravity is propagated by a massive field, i.e. by a graviton with some small, nonzero mass $m_g$ For detailed reviews on massive gravity including theories without ghosts see \cite{rham14,rham17,babi09} and references therein (earlier, a presence of Boulware -- Deser \cite{boul72} ghosts was considered as a significant pathology of massive gravity theories). Ever since they were first introduced by Fierz and Pauli in 1939 \cite{fier39}, such so called theories of massive gravity have gained a significant attention due to their ability to provide a possible explanation for the accelerated expansion of the Universe without dark energy (DE) hypothesis, and due to important predictions that the velocity of gravitational waves (gravitons) should depend on their frequency, as well as that the effective gravitational potential should include a nonlinear (exponential) correction of Yukawa form, depending on the Compton wavelength of graviton: $\lambda_g=h/(m_g\,c)$. Namely, if gravitation is propagated by a massive field, then the effective Newtonian potential has a Yukawa form: $\propto r^{-1}\exp(-r/\lambda_g)$, and the massive graviton then propagates at an energy $E$ (or frequency $f$) dependent speed: $v^2_g/c^2=1 - m_g^2c^4/E^2=1-h^2c^2/(\lambda_g^2E^2)= 1 - c^2/(f\lambda_g)^2$ \cite{will98,will14}. The above modified dispersion relation is obtained from the modified special relativistic relation between energy $E$ and momentum $p$ of graviton: $E^2 = p^2 c^2 + m_g^2c^4$, and by taking that its velocity $v_g$ satisfies: $v^2_g/c^2 \equiv c^2 p^2/E^2$. Moreover, one can obtain the observational constraints on the $v_g/c$ ratio from the time difference $\Delta t \equiv \Delta t_\mathrm{a} - (1+z) \Delta t_\mathrm{e}$ between a gravitational wave and an electromagnetic wave from the same event, where $\Delta t_\mathrm{a}$ and $\Delta t_\mathrm{e}$ are the differences in arrival time and emission time, respectively, of these two signals, and $z$ is the redshift of the source \cite{will98}. The fractional speed difference can be then obtained using the following expression: $1-\dfrac{v_\mathrm{g}}{c}=5\times 10^{-17}\left(\dfrac{200\mathrm{\ Mpc}}{D}\right)\left(\dfrac{\Delta t}{1\mathrm{\ s}}\right)$, where $D$ is distance of the source \cite{will98,will14}.

Below we present a short overview of observational tests of massive gravity theories, and their applications for obtaining the constraints on Compton wavelength and mass of graviton, as well as on speed of gravity (for more detailed review see e.g. \cite{rham17}).

Yukawa-like potential of the form $U\left(r\right)=\dfrac{G_{\infty}M}{r}\left(1+\alpha\,e^{\displaystyle -r/r_0}\right)$ was studied in \cite{sand84}, where $\alpha$ is the strength of Yukawa interaction, $r_0$ is a characteristic length scale, while the gravitational constant measured locally ($G_0$) and at infinity ($G_\infty$) are related by: $G_0=G_\infty\left(1+\alpha\right)$. It was found that, if the length scale $r_0$ corresponds to graviton mass $m_0$ as $r_0=\dfrac{h}{m_0 c}$, then the flat rotation curves of spiral galaxies could be accounted for $\alpha\sim -1$ and without introducing dark matter (DM) hypothesis \cite{sand84}. The negative sign of the strength of Yukawa interaction $\alpha$ indicated an additional repulsive (anti-gravity) force which could mimic the effects of dark energy on the large scales. Thus, it was demonstrated that this type of correction on gravitational potential could be used to account for both DM at galactic and DE at cosmological scales, which is one of the main reasons while Yukawa gravity has recently attracted the significant attention.

Different experimental tests and constraints on the range and strength of additional Yukawa gravitational interaction between masses $m_1$ and $m_2$ in the form $V\left(r\right)=-G_N\dfrac{m_1\,m_2}{r}\cdot\left(1+\alpha\,e^{\displaystyle-r/\lambda}\right)$ were reviewed in \cite{adel09}. These constraints covered a wide range of length scales, including short-range laboratory experiments with torsion balance, lunar laser ranging (LLR) and Solar System tests, and were summarized in Figs. 9 and 10 of \cite{adel09}. In the case of LLR and Solar System, the following constraints on the range of Yukawa interaction $\Lambda$ were obtained, respectively: $\lambda \gg 4 \times 10^8\ \mathrm{m}$ and $\lambda \gg 1.5 \times 10^{11}\ \mathrm{m}$.

LIGO Scientific and Virgo Collaborations used the following Yukawa type correction to the gravitational potential with characteristic length scale $\lambda_g$: $\varphi\left(r\right)=\dfrac{GM}{r}\left(1-e^{\displaystyle -r/\lambda_g}\right)$ in order to discuss and compare the so-called static and dynamical bounds on graviton mass \cite{abbo16a}. In the first publication on the discovery of gravitational waves and binary black holes the authors found a graviton mass constraint from a dispersion relation therefore, a massive gravity theory was considered as feasible alternative for GR. The static bounds, such as those from the Solar System observations, do not probe the propagation of gravitational interaction, in contrast to dynamical bounds which are obtained by studying the very massive, compact objects such as binary pulsars and supermassive Kerr black holes. Fig. 8 of \cite{abbo16} shows a comparison of cumulative posterior probability distribution and exclusion regions for the graviton Compton wavelength, obtained from the first gravitational wave signal GW150914 with exclusion regions obtained from the binary pulsar J0737-3039 and the Solar System observations. The resulting dynamical bound of $\lambda_g>10^{13}\;\mathrm{km}$ at 90\% confidence, which corresponds to a graviton mass $m_g\le 1.2\times 10^{-22}\;\mathrm{eV}$, obtained from GW150914, was approximately for a factor of three better than the existing Solar-System static bounds \cite{abbo16} (after data analysis of the third working run LIGO--Virgo--KAGRA consortium improved previous estimates as $m_g\le 1.27\times 10^{-23}\;\mathrm{eV}$ \cite{abbo21}). This LIGO--Virgo bound on graviton mass is currently considered as the most robust, although some other techniques could in principle provide more stringent constraints. For example, expected detection limit for a future pulsar timing array (PTA) with 300 pulsars, observed for 10 years is $m_g=5\times 10^{-23}\;\mathrm{eV}$ \cite{lee10} (however, here we should emphasize that we compare the found graviton mass bounds from gravitational wave observations and constraints which will be expected to find from future pulsar timing observations), while the current studies of weak lensing, which are model-dependent and very sensitive to the uncertainties on the amount and spatial distribution of DM in the Universe, provide the limit of $\lambda_g>6.82\;\mathrm{Mpc}$ which corresponds to $m_g<6\times 10^{-30}\;\mathrm{eV}$ \cite{rana18}.

However, further progress in tightening the bound on $\lambda_g$ (and thus on the graviton mass) was recently achieved using Solar system data on the perihelion advance of planets from Mercury to Saturn, obtained by high-precision radar tracking of planets and spacecrafts, as well as using improved ephemeris computer codes \cite{will18}. Assuming the Newtonian gravitational potential given by the Yukawa form $(Gm/r) e^{-r/\lambda_g}$, the following approximate expression (up to leading order in $a/\lambda_g$ ratio) for the perihelion advance per orbit was obtained: $\Delta\varpi\approx\pi\left(\dfrac{a}{\lambda_g}\right)^2 (1-e^2)^{-1/2}$, where $\varpi$ is the longitude of perihelion, $a$ is the orbital semimajor axis and $e$ is the orbital eccentricity \cite{will18}. The resulting lower bound on $\lambda_g$ was between $1.2$ and $2.2 \times 10^{14}\ \mathrm{km}$ (see Table 1 in \cite{will18}), surpassing the first LIGO/Virgo graviton bounds by an order of magnitude.

Recently, there has been also an attempt to constrain the speed of gravity from the observed time difference between gravitational wave event GW170817, observed by the Advanced LIGO and Virgo detectors, and $\gamma$-ray burst  GRB170817A, observed independently by $\gamma$-ray detectors onboard the Fermi and the International Gamma-Ray Astrophysics Laboratory (INTEGRAL) space observatories \cite{abbo17}. These two GW and $\gamma$-ray signals were emitted from a binary neutron star merger in the nearby galaxy NGC 4993, located at redshift $z\approx 0.01$, which corresponds to the luminosity distance of $D = 26\ \mathrm{Mpc}$. Optical and IR telescopes (Swope, DLT40, MASTER, DECam, VISTA et al.) played  a crucial for the localization of the event and a power of multi-messenger astronomy was remarkably demonstrated \cite{abbo17a}. The observed time difference between GRB 170817A and GW170817 was $1.74\pm 0.05\ \mathrm{s}$, while their emitted time difference was unknown \cite{abbo17}. Therefore, the upper and lower bounds on speed of gravity were estimated assuming that the peak of the gravitational wave signal and the first $\gamma$-photons were emitted simultaneously (i.e. by attributing the entire observed lag to the faster gravitational wave signal), and that $\gamma$-ray signal was emitted 10 s after the gravitational wave signal, respectively. This resulted with the following constraint on the fractional speed difference: $-3\times 10^{-15}\le\dfrac{v_\mathrm{g}}{c}-1\le +7\times 10^{-16}$ \cite{abbo17}.

This research is continuation of our previous investigations in this field  \cite{bork13,capo14,jova21,zakh16a,zakh16b,zakh17a,zakh17b,zakh18a,zakh18b}, in which we constrained the parameters of Yukawa gravity using the observed stellar orbits around the central SMBH of the Milky Way  \cite{ghez08,gill09a,gill09b,gill17,hees17,chu18,amor19,hees20}, as well as developed a novel and independent method for obtaining the graviton mass bounds from these observations which are, since 2019, accepted by the Particle Data Group (PDG) and included in their ''Gauge and Higgs Boson Particle Listings'' \cite{zyla20}. In addition, here we will extend our study on the possible influence of the strength of Yukawa interaction, described by the parameter $\delta$ in the gravitational potential, on the obtained bounds on graviton mass. This paper is organized as follows: in \S 2 we present the procedure for obtaining the gravitational potential with a Yukawa-like correction in the Newtonian limit of any analytic $f(R)$ gravity model, in \S 3 we describe some basic properties of S-stars and their observed orbits around the SMBH at GC, as well as our method for obtaining the Compton wavelength and graviton mass from analysis of the observed stellar orbits in Yukawa gravity, in \S 4 we discuss the main results of our study, especially regarding the strength of Yukawa interaction $\delta$, and finally in \S 5 we point out of our most important conclusions.

\section{Yukawa-like nonlinear correction to the gravitational potential}

Yukawa-like potentials are characterized by presence of decreasing exponential terms and thats why deviate from the standard Newtonian gravitational potential \cite{talm88,sere06,card11,sand84,iori07,iori08}.

The Yukawa term is analysed in case of short and long ranges. The Yukawa term for the short ranges is analysed in \cite{adel09} and references therein, and for the longer ranges the parameters of Yukawa gravity potential are given for clusters of galaxies \cite{capo07b,capo09b}, for rotation curves of spiral galaxies \cite{card11} and for the binary pulsars \cite{miao19,dong22}. Also, other studies of long-range Yukawa term investigations can be found in \cite{whit01,amen04,reyn05,seal05,sere06,moff05,moff06}.

The action of Yukawa-like nonlinear correction to the gravitational potential in the Newtonian limit can be given in the form \cite{capo14}:
\begin{equation}
{\cal S}= \int  d^{4}x\sqrt{-g}\left[f(R)+\mathcal{X}\mathcal{L}_m\right],\quad \mathcal{X}=\dfrac{16\pi G}{c^4},
\label{equ2.1}
\end{equation}

\noindent where $f$ is a generic function of Ricci scalar curvature $R$ and $\mathcal{X}$ is the coupling constant.

The resulting $4^\mathrm{th}$-order field equations are following:
\begin{equation}
f'(R)R_{\mu\nu}-\frac{1}{2}f(R)g_{\mu\nu}-f'(R)_{;\mu\nu}+g_{\mu\nu}\Box f'(R)=\frac{\mathcal{X}}{2}T_{\mu\nu}.
\label{equ2.2}
\end{equation}

Trace of these equations is given by:

\begin{equation}
3\Box f'(R)+f'(R)R-2f(R)=\frac{\mathcal{X}}{2}T.
\label{equ2.3}
\end{equation}

Expansion of analytic Taylor expandable $f(R)$ functions with respect to the value $R = 0$ (i.e. around the Minkowskian background) has the following form:

\begin{equation}
f(R)\,=\,\sum_{n\,=\,0}^{\infty}\frac{f^{(n)}(0)}{n!}\,R^n\,=\,f_0+f_1R+\frac{f_2}{2}R^2+\ldots
\label{equ2.4}
\end{equation}

By adopting the spherical symmetry, the metric can be recast as follows \cite{stab13}:
\begin{equation}
ds^2=\left[1+\frac{2\Phi(r)}{c^2}\right]c^2dt^2-\left[1-\frac{2\Psi(r)}{c^2}\right]dr^2-r^2d\Omega^2,
\label{equ2.5}
\end{equation}

\noindent where $\Phi\left(r\right)$ and $\Psi\left(r\right)$ are two potentials given by \cite{capo14,stab13}:

\begin{equation}
\Phi\left(r\right)=-\dfrac{GM}{(1+\delta)r}\left({1+\delta e^{-\dfrac{r}{\Lambda}}}\right),
\label{equ2.6}
\end{equation}

\begin{equation}
\Psi\left(r\right)=\dfrac{G M}{(1+\delta)r}\biggl[\left(1+\dfrac{r}{\Lambda}\right)\delta e^{-\dfrac{r}{\Lambda}}-1\biggr].
\label{equ2.7}
\end{equation}

The parameter $\Lambda$ is the range of interaction ($\Lambda^2=-f_1/f_2$) and depends on the typical scale of a gravitational system. The second parameter $\delta$ is the strength of interaction ($\delta=f_1-1$).

In our previous investigations we constrained different Extended Gravity theories using astronomical data for different astrophysical systems: the S2 star orbit \cite{bork12,bork13,zakh14,bork16,zakh16a,zakh18a,dial19,bork19,jova21}, fundamental plane of elliptical galaxies \cite{bork16a,capo20,bork21} and baryonic Tully-Fischer relation of spiral galaxies \cite{capo17}. Here, we analyse stellar orbits (so called S-stars) around the central SMBH of the Milky Way in the frame of Yukawa gravity in order to constrain gravity parameters \cite{bork13,zakh16a,capo14,jova21,zakh16b,zakh17a,zakh17b,zakh18a,zakh18b}.

\section{Analysis of the stellar orbits around Sgr A* in Yukawa gravity}

S-stars are the bright stars which move around the GC \cite{ghez00,scho02,ghez08,gill09a,gill09b,genz10,meye12,gill17,hees17,chu18,amor19,hees20,abut20} where a compact bright radio source Sgr A$^\ast$ is located. The GC consists of the SMBH with mass around $4.3\times 10^6 M_\odot$ and an extended mass distribution formed with stellar cluster, interstellar and probably dark matter. We assume that total mass of bulk distribution inside a spherical shell where trajectories of bright stars are located is much smaller than the BH mass and recent estimates done by the GRAVITY collaboration support these assumptions since for feasible density profiles the total extended mass inside a ball with the S2 apocenter radius must be less than $3000~M_\odot$ at $1~\sigma$ confidence level \cite{abut22}. Also, using observations in May 2018, GRAVITY Collaboration \cite{abut18,abut19} and the Keck group \cite{do19} evaluated relativistic redshifts of spectral lines for the S2 star near its periapsis passage. The obtained results showed that the redshifts were consistent with theoretical estimates done in the first post-Newtonian correction of GR. The orbits of S-stars around Sgr A$^\ast$ are monitored for about 30 years by the following telescopes: New Technology Telescope and Very Large Telescope (NTT/VLT) in Chile \cite{gill09a,gill09b} (recently VLT units started to act as the GRAVITY  interferometer) and by Keck telescopes in Hawaii \cite{ghez08}. These two groups performed the precise astrometric observations of S2 star. Nowadays, there are more recent and precise observations not only of S2 star, but also of several other members of the S-star cluster, such as S38, S55 and S62 \cite{gill17,pars17,abut18,peis20} and thus it is possible to perform the orbital analysis of these stars (see e.g. \cite{kali20,lalr21,lalr22,doku15,mart18,dela18,dadd21,bork21b,bork22,beni22,gain20}).

We assume that S-stars move around the central SMBH of our Galaxy at a sufficiently large distances where the space-time is practically flat, and we studied the PPN equations of motion in both Yukawa gravity and GR. Besides the well known first post-Newtonian correction of GR, PPN equations of motion in Yukawa gravity also include an additional exponential correction which arises because the $f(R)$ theories of gravity with Yukawa-like potentials do not entirely fit into the standard PPN formalism and require its extension/modification (see e.g. \cite{clif08,alsi12} and references therein). Therefore, in this modified PPN formalism the equations of motion in Yukawa gravity have the following form:
 
\begin{equation}
\vec{\ddot{r}}_{\scriptscriptstyle Y} = \vec{\ddot{r}}_{\scriptscriptstyle N} + \vec{\ddot{r}}_{cor,\scriptscriptstyle PPN} + \vec{\ddot{r}}_{cor,\scriptscriptstyle Y},
\label{equ3.1}
\end{equation}

\noindent where the $\vec{\ddot{r}}_{\scriptscriptstyle N}$ is the Newtonian acceleration, $\vec{\ddot{r}}_{cor,\scriptscriptstyle PPN}$ is its first post-Newtonian correction in the PPN formalism and $\vec{\ddot{r}}_{cor,\scriptscriptstyle Y}$ is additional Yukawa correction. These three contributions are given by the following expressions (see Eq. (2.1) from Ref. \cite{dong22}):

\begin{eqnarray}
\vec{\ddot{r}}_{\scriptscriptstyle N} &=& -GM \dfrac{\vec{r}}{r^3} \nonumber\\
\vec{\ddot{r}}_{cor,\scriptscriptstyle PPN} &=& \dfrac{GM}{c^2r^3} \Bigg\{\left[2\left(\beta+\gamma\right)\dfrac{G M}{r}-\gamma\left(\vec{\dot{r}}\cdot\vec{\dot{r}}\right)\right] \vec{r} + 2\left(1+\gamma\right)\left(\vec{r}\cdot\vec{\dot{r}}\right)\vec{\dot{r}} \Bigg\} \nonumber\\
\vec{\ddot{r}}_{cor,\scriptscriptstyle Y} &=& \dfrac{\delta \cdot GM}{1 + \delta} \left[ 1 - \left(1 - \dfrac{r}{\Lambda} \right) e^{-\dfrac{r}{\Lambda}} \right] \dfrac{\vec{r}}{r^3}.
\label{equ3.2}
\end{eqnarray}

The standard PPN equations of motion in the GR case are given by:

\begin{equation}
\vec{\ddot{r}}_{\scriptscriptstyle GR} = \vec{\ddot{r}}_{\scriptscriptstyle N} + \vec{\ddot{r}}_{cor,\scriptscriptstyle PPN}.
\label{equ3.3}
\end{equation} 

Simulated orbits of S-stars in Yukawa gravity and GR were then obtained by numerical integration of the expressions (\ref{equ3.1}) and (\ref{equ3.3}), respectively. Both $\beta$ and $\gamma$ PPN parameters in $\vec{\ddot{r}}_{cor,\scriptscriptstyle PPN}$ are taken to be equal to 1, since Yukawa gravity is indistinguishable from GR up to the first post-Newtonian correction \cite{clif08}. The obtained results are used for studying the orbital precession caused by Yukawa type correction in the gravitational potential and mutual comparisons between the orbits simulated in Yukawa gravity and in GR.

In this paper we also assumed that the orbital precession of the S-stars is close to the corresponding GR prediction and only slightly deviates from it. This assumption was based on the fact that the GRAVITY Collaboration detected the orbital precession of the S2 star around the SMBH and showed that it was close to the corresponding prediction of GR \cite{abut20}. Taking this into account, we derived expression for parameter $\Lambda$ under an assumption that the orbital precession of S-stars in Yukawa gravity deviates from the Schwarzschild precession in GR by a specific factor $f_{SP} = 1.10$, as indicated in \cite{abut20}.

The approximate formula for the additional contribution of Yukawa gravity to the Schwarzschild precession for $a\ll\Lambda$ is \cite{zakh16a}:

\begin{equation}
\Delta\varphi_Y^{rad}\approx\dfrac{\pi\delta\sqrt{1-e^2}}{1+\delta}\dfrac{a^2}{\Lambda^2},
\label{equ3.4}
\end{equation}

\noindent where $e$ is orbital eccentricity and $a$ is the semi-major axis of the orbital ellipse. The total orbital precession can be obtained by adding the above expression to the formula for Schwarzschild precession \cite{will14}:

\begin{equation}
\Delta\varphi_{GR}^{rad}\approx\dfrac{6\pi G M}{c^2 a(1-e^2)},
\label{equ3.5}
\end{equation}

\noindent and it is expected to be close to the observed precession:

\begin{equation}
\Delta\varphi_Y + \Delta\varphi_{GR} \approx \Delta\varphi_{obs},
\label{equ3.6}
\end{equation}
\noindent where $\Delta\varphi_Y$, $\Delta\varphi_{GR}$ and $\Delta\varphi_{obs}$ are Yukawa contribution, GR contribution and observed precession \cite{abut20}, respectively. This can be recast as:

\begin{equation}
\dfrac{2\delta \sqrt{1 - e^2}}{1 + \delta} \dfrac{a^2}{\Lambda^2} + \dfrac{6\pi GM}{c^2 a(1 - e^2)} \approx \dfrac{2\pi GM}{c^2 a(1 - e^2)} \left(3 f_{sp}\right),
\label{equ3.7}
\end{equation}

\noindent where parameter $f_{sp}$ is defined as: $f_{sp} = \dfrac{2 + 2\gamma - \beta}{3}$, and it characterizes how relativistic the model is. The value of $f_{sp} = 1.10 \pm 0.19$ was measured by GRAVITY Collaboration \cite{abut20}. Using the third Kepler law (since the orbits are very close to Newtonian ones):

\begin{equation}
\dfrac{P^2}{a^3} \approx \dfrac{4\pi^2}{GM},
\label{equ3.8}
\end{equation}

\noindent it was found that, according to the expression (\ref{equ3.7}), $\Lambda$ has to satisfy the following condition:

\begin{equation}
\Lambda(P,e;\delta) \approx \dfrac{cP}{2\pi} \sqrt{\dfrac{\delta (\sqrt{1-e^2})^3}{2(3f_{sp} - 3)(1 + \delta)}}.
\label{equ3.9}
\end{equation}

We can see that for the given values of universal constant $\delta$ and parameter $f_{sp}$, the following condition holds: $\Lambda\propto c\,P^{*}$, where $P^{*} = P(1-e^2)^{3/4}$. Thus, $P^{*}$ can be used as a criterion for classification of the gravitational systems according to their scales, as well as to find the systems which could be described by similar values of the range of Yukawa interaction $\Lambda$.

\section{Results and discussion}

\subsection{Constrains on Yukawa gravity parameters from the stellar orbits around GC}

In order to study the influence of strength of Yukawa interaction $\delta$ on its range $\Lambda$ in the case of different S-stars, we estimated the values of $\Lambda$ and its absolute errors for the following three values of $\delta$: $\delta=0.1, \delta=10$ and $\delta=100$. The obtained estimates are given in Tables \ref{tab1} and \ref{tab2} in which the following values are given for each S-star: orbital precession ($\Delta\varphi$), scale criterion $P^{*} = P\,(1-e^2)^{3/4}$, range of Yukawa interaction ($\Lambda$) and its absolute error ($\Delta\Lambda$). The observed orbital elements and their uncertainties are taken from Table 3 of \cite{gill17} for all given S-stars, except of S111. It can be seen from Tables \ref{tab1} and \ref{tab2} that universal constant $\delta$ strongly affects the range of Yukawa interaction $\Lambda$ and that they are correlated in the case of weaker Yukawa interaction (i.e. for smaller values of $\delta$, such as $\delta$ = 0.1).

\begin{figure}[ht!]
\centering
\includegraphics[width=0.53\textwidth]{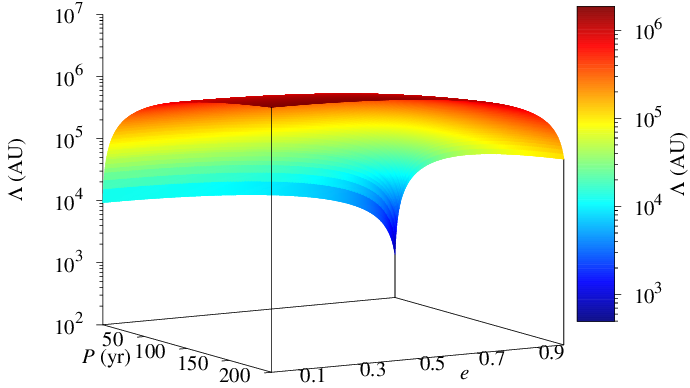}\hfill
\includegraphics[width=0.46\textwidth]{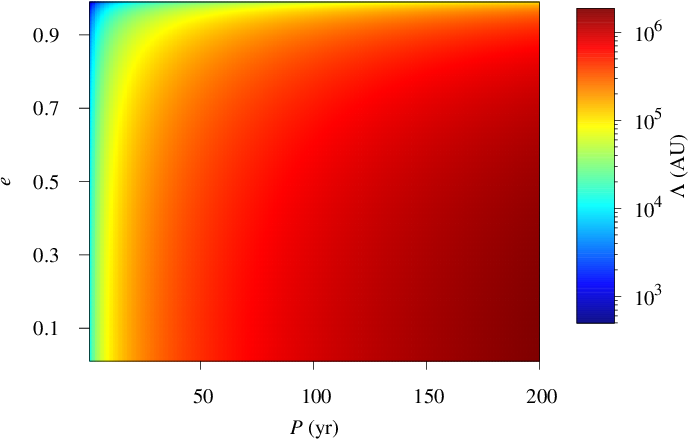}\vspace*{0.5cm}
\includegraphics[width=0.49\textwidth]{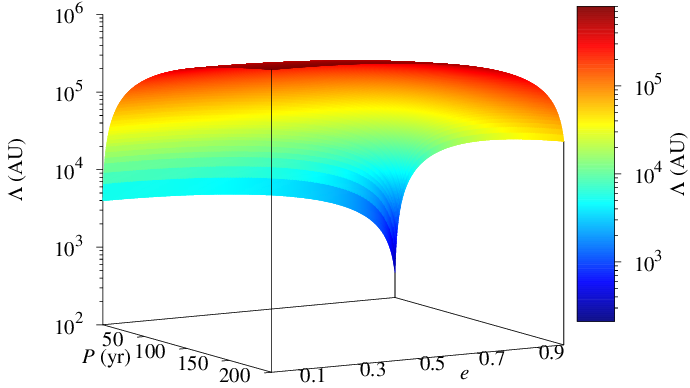}\hfill
\includegraphics[width=0.49\textwidth]{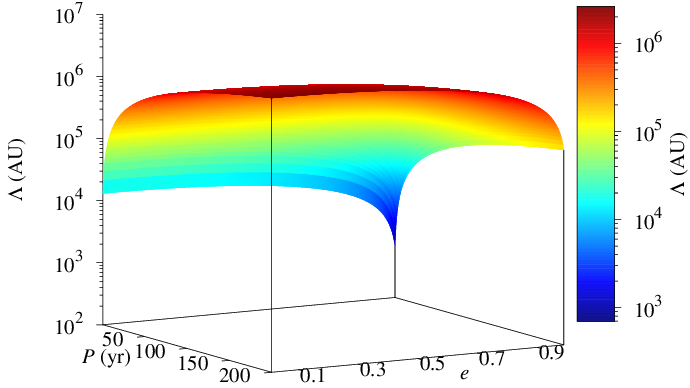}
\caption{The 3D graphic (top left) of the function $\Lambda\left(P,e\right)$ defined in (\ref{equ3.6}) and representing the dependence of the range of Yukawa interaction $\Lambda$ on the orbital period $P$ (yr) and eccentricity $e$ of S-stars for $\delta = 1$, as well as its projection to the $P-e$ parameter space (top right). Bottom panels  shows the same as in top left panel, but for $\delta = 0.1$ (bottom left) and $\delta = 100$ (bottom right).}
\label{fig01}
\end{figure}

\begin{figure}[ht!]
\centering
\includegraphics[width=0.5\textwidth]{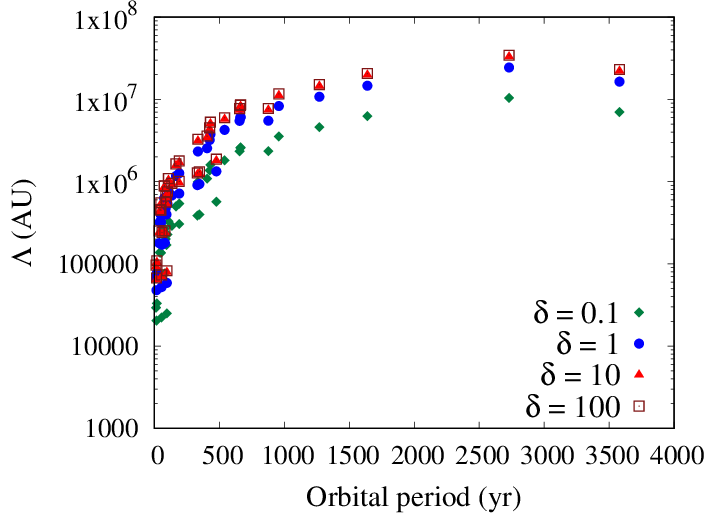}\hfill
\includegraphics[width=0.5\textwidth]{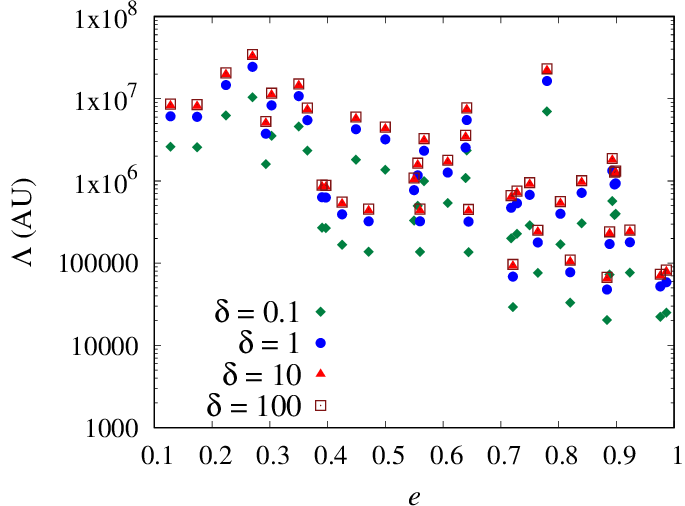}
\caption{The graphic of range of interaction $\Lambda$ vs. observed orbital period $P$ (left) and eccentricity $e$ (right) of S-stars for 4 different values of parameter $\delta$ equals 0.1, 1, 10 and 100, respectively.}
\label{fig02}
\end{figure}

Effects of 4 different values of parameter $\delta$ (0.1, 1, 10 and 100) on estimates of parameter $\Lambda$, are presented in Figs. \ref{fig01}-\ref{fig02}. Several 3D graphics of the function $\Lambda\left(P,e\right)$, representing $\Lambda$ dependence on orbital period $P$ (yr) and eccentricity $e$ of S-stars, are shown in Fig. \ref{fig01} (see also Tables \ref{tab1} and \ref{tab2} for some particular S-stars). Top panels of Fig. \ref{fig01} show the case for $\delta = 1$, while the bottom panels present two cases: $\delta = 0.1$ (bottom left) and $\delta = 100$ (bottom right). From Fig. \ref{fig01} it is evident that the larger values for the $\Lambda$ are obtained for the S-stars with larger orbital periods $P$ and lower eccentricities $e$. The shape of surface which represents the values of the $\Lambda$ is similar for all studied values of $\delta$, but the values of $\Lambda$ are different for different values of Yukawa strengths $\delta$. In the case of $\delta=1$, the values for $\Lambda$ are ranging from $10^{4}$ to $10^{6}$ AU (see Fig. \ref{fig01}). Especially, this difference is noticeable if one compares the case for $\delta=0.1$ with other studied cases.

\begin{table}
\centering
\small
\caption{Orbital precession ($\Delta\varphi$), scale criterion ($P^{*}$) for S-stars, range of Yukawa interaction ($\Lambda$) and its absolute errors ($\Delta\Lambda$). The range of Yukawa interaction and its absolute errors are estimated for the following three values of $\delta$: $\delta=0.01, \delta=0.1$ and $\delta=1$. The observed orbital elements and their uncertainties are taken from Table 3 of \cite{gill17} for all given S-stars, except of S111.}
\setlength{\tabcolsep}{2pt}
\begin{tabular}{|l|r|r|rcl|rcl|rcl|}
\hline
\multicolumn{1}{|l|}{Star}&\multicolumn{1}{c|}{$\Delta\varphi$}&\multicolumn{1}{c|}{$\ P^{*}$}&\multicolumn{9}{c|}{$\Lambda\pm\Delta\Lambda$ (AU)} \\
\cline{4-12}
&\multicolumn{1}{c|}{('')}&\multicolumn{1}{c|}{(yr)}&\multicolumn{3}{c|}{$\delta=0.01$}&\multicolumn{3}{c|}{$\delta=0.1$}&\multicolumn{3}{c|}{$\delta=1$} \\
\hline
\hline
S1   &   48.2 &  125.79 &  164626.9 & $\pm$ &    13537.8 &   498844.4 & $\pm$ &    41021.7 &  1169893.8 & $\pm$ &    96204.5 \\
S2   &  722.1 &  5.12   &    6730.8 & $\pm$ &      149.9 &    20395.2 & $\pm$ &      454.3 &    47831.0 & $\pm$ &     1065.4 \\
S4   &   65.5 &  68.01  &   89185.0 & $\pm$ &     1750.2 &   270244.0 & $\pm$ &     5303.3 &   633778.4 & $\pm$ &    12437.2 \\
S6   &  102.4 &  76.74  &  100839.1 & $\pm$ &      267.5 &   305557.6 & $\pm$ &      810.6 &   716596.2 & $\pm$ &     1901.1 \\
S8   &  138.0 &  42.73  &   56050.5 & $\pm$ &     1717.2 &   169841.6 & $\pm$ &     5203.5 &   398314.0 & $\pm$ &    12203.2 \\
S9   &  124.3 & 34.33   &   45030.4 & $\pm$ &     2503.1 &   136448.8 & $\pm$ &     7584.9 &   320000.8 & $\pm$ &    17788.2 \\
S12  &  314.6 &  18.33  &   24050.7 & $\pm$ &      475.7 &    72877.1 & $\pm$ &     1441.4 &   170912.0 & $\pm$ &     3380.4 \\
S13  &   91.6 &  42.20  &   55328.1 & $\pm$ &      601.8 &   167652.4 & $\pm$ &     1823.5 &   393179.8 & $\pm$ &     4276.6 \\
S14  & 1465.9 &  5.60   &    7346.3 & $\pm$ &      981.2 &    22260.4 & $\pm$ &     2973.2 &    52205.2 & $\pm$ &     6972.8 \\
S17  &   66.1 &  67.35  &   88370.7 & $\pm$ &     4262.7 &   267776.5 & $\pm$ &    12916.7 &   627991.6 & $\pm$ &    30292.4 \\
S18  &  107.1 &  34.71  &   45506.5 & $\pm$ &      926.2 &   137891.7 & $\pm$ &     2806.5 &   323384.7 & $\pm$ &     6581.8 \\
S19  &   87.1 &  72.62  &   95481.5 & $\pm$ &    36447.7 &   289323.6 & $\pm$ &   110442.1 &   678523.9 & $\pm$ &   259009.7 \\
S21  &  217.4 &  19.18  &   25142.3 & $\pm$ &     1261.7 &    76185.0 & $\pm$ &     3823.2 &   178669.6 & $\pm$ &     8966.2 \\
S22  &   19.0 &  456.10 &  599449.8 & $\pm$ &   236689.3 &  1816423.7 & $\pm$ &   717204.5 &  4259891.2 & $\pm$ &  1681993.5 \\
S23  &  114.1 &  34.54  &   45425.0 & $\pm$ &    11014.4 &   137644.5 & $\pm$ &    33375.3 &   322805.0 & $\pm$ &    78272.1 \\
S24  &  107.5 &  97.28  &  127590.2 & $\pm$ &    14036.6 &   386617.6 & $\pm$ &    42533.2 &   906698.7 & $\pm$ &    99749.1 \\
S29  &   98.5 &  57.33  &   75236.9 & $\pm$ &    14099.5 &   227979.3 & $\pm$ &    42723.7 &   534658.8 & $\pm$ &   100196.0 \\
S31  &   63.3 &  82.46  &  108733.3 & $\pm$ &     3953.7 &   329478.4 & $\pm$ &    11980.3 &   772695.3 & $\pm$ &    28096.4 \\
S33  &   47.9 &  135.83 &  178323.3 & $\pm$ &    27097.8 &   540346.5 & $\pm$ &    82110.5 &  1267224.9 & $\pm$ &   192566.2 \\
S38  &  427.5 &  8.31   &   10917.4 & $\pm$ &       51.8 &    33081.5 & $\pm$ &      157.1 &    77582.9 & $\pm$ &      368.4 \\
S39  &  364.5 &  19.25  &   25286.5 & $\pm$ &     2038.3 &    76622.1 & $\pm$ &     6176.3 &   179694.7 & $\pm$ &    14484.7 \\
S42  &   30.8 &  250.45 &  327668.0 & $\pm$ &   127216.9 &   992883.5 & $\pm$ &   385486.4 &  2328518.3 & $\pm$ &   904045.8 \\
S54  &   81.5 &  144.02 &  187868.3 & $\pm$ &   301214.1 &   569269.5 & $\pm$ &   912724.3 &  1335055.3 & $\pm$ &  2140528.3 \\
S55  &  382.8 &  7.38   &    9666.0 & $\pm$ &      302.1 &    29289.4 & $\pm$ &      915.3 &    68689.7 & $\pm$ &     2146.6 \\
S60  &  105.5 &  50.59  &   66370.5 & $\pm$ &     2549.6 &   201112.8 & $\pm$ &     7725.8 &   471651.3 & $\pm$ &    18118.6 \\
S66  &   13.4 &  655.82 &  860607.2 & $\pm$ &    88872.3 &  2607770.2 & $\pm$ &   269296.7 &  6115763.2 & $\pm$ &   631556.7 \\
S67  &   19.3 &  402.94 &  528751.9 & $\pm$ &    32803.8 &  1602198.3 & $\pm$ &    99400.4 &  3757488.1 & $\pm$ &   233114.5 \\
S71  &  106.2 &  100.28 &  131669.4 & $\pm$ &    20154.0 &   398978.3 & $\pm$ &    61069.7 &   935687.0 & $\pm$ &   143221.2 \\
S83  &   15.3 &  589.30 &  773374.9 & $\pm$ &   184565.2 &  2343443.1 & $\pm$ &   559260.5 &  5495861.3 & $\pm$ &  1311582.1 \\
S85  &   11.0 & 1772.21 & 2311796.7 & $\pm$ &  3523752.4 &  7005094.4 & $\pm$ & 10677503.5 & 16428402.6 & $\pm$ & 25040965.4 \\
S87  &    7.6 & 1577.89 & 2065566.0 & $\pm$ &   200654.2 &  6258978.1 & $\pm$ &   608012.7 & 14678604.7 & $\pm$ &  1425916.1 \\
S89  &   31.0 & 273.90  &  358908.7 & $\pm$ &    49485.2 &  1087547.9 & $\pm$ &   149947.8 &  2550525.9 & $\pm$ &   351658.7 \\
S91  &   11.4 & 891.25  & 1168818.2 & $\pm$ &   101284.2 &  3541696.1 & $\pm$ &   306906.4 &  8306013.7 & $\pm$ &   719759.3 \\
S96  &   13.6 & 646.91  &  848925.2 & $\pm$ &    53447.7 &  2572372.1 & $\pm$ &   161954.7 &  6032747.3 & $\pm$ &   379817.5 \\
S97  &    9.7 & 1151.43 & 1516520.2 & $\pm$ &   550839.2 &  4595286.1 & $\pm$ &  1669126.3 & 10776901.1 & $\pm$ &  3914448.1 \\
S145 &   23.6 & 343.33  &  452172.2 & $\pm$ &   222048.9 &  1370150.3 & $\pm$ &   672841.7 &  3213287.3 & $\pm$ &  1577953.6 \\
S175 & 1812.0 & 6.30    &    8263.6 & $\pm$ &     2000.9 &    25040.0 & $\pm$ &     6062.9 &    58723.9 & $\pm$ &    14218.7 \\
R34  &   18.6 & 589.73  &  775088.0 & $\pm$ &   220322.6 &  2348634.1 & $\pm$ &   667611.0 &  5508035.2 & $\pm$ &  1565686.6 \\
R44  &    5.5 & 2579.33 & 3444472.6 & $\pm$ &  2260986.2 & 10437273.8 & $\pm$ &  6851130.7 & 24477576.8 & $\pm$ & 16067325.7 \\
\hline
\end{tabular}
\normalsize
\label{tab1}
\end{table}

\begin{table}
\centering
\small
\caption{The same as Table \ref{tab1}, but for the following three values of $\delta$: $\delta=10, \delta=100$ and $\delta=1000$.}
\setlength{\tabcolsep}{2pt}
\begin{tabular}{|l|r|r|rcl|rcl|rcl|}
\hline
\multicolumn{1}{|l|}{Star}&\multicolumn{1}{c|}{$\Delta\varphi$}&\multicolumn{1}{c|}{$\ P^{*}$}&\multicolumn{9}{c|}{$\Lambda\pm\Delta\Lambda$ (AU)} \\
\cline{4-12}
&\multicolumn{1}{c|}{('')}&\multicolumn{1}{c|}{(yr)}&\multicolumn{3}{c|}{$\delta=10$}&\multicolumn{3}{c|}{$\delta=100$}&\multicolumn{3}{c|}{$\delta=1000$} \\
\hline
\hline
S1   &   48.2 &  125.79 &  1577484.5 & $\pm$ &   129722.1 &  1646268.8 & $\pm$ &   135378.5 &  1653653.1 & $\pm$ &   135985.7 \\
S2   &  722.1 &  5.12   &    64495.3 & $\pm$ &     1436.6 &    67307.6 & $\pm$ &     1499.2 &    67609.5 & $\pm$ &     1506.0 \\
S4   &   65.5 &  68.01  &   854586.6 & $\pm$ &    16770.4 &   891849.8 & $\pm$ &    17501.6 &   895850.2 & $\pm$ &    17580.1 \\
S6   &  102.4 &  76.74  &   966258.1 & $\pm$ &     2563.5 &  1008390.6 & $\pm$ &     2675.3 &  1012913.7 & $\pm$ &     2687.3 \\
S8   &  138.0 &  42.73  &   537086.4 & $\pm$ &    16454.8 &   560505.5 & $\pm$ &    17172.3 &   563019.6 & $\pm$ &    17249.3 \\
S9   &  124.3 & 34.33   &   431489.0 & $\pm$ &    23985.5 &   450303.6 & $\pm$ &    25031.4 &   452323.4 & $\pm$ &    25143.7 \\
S12  &  314.6 &  18.33  &   230457.7 & $\pm$ &     4558.1 &   240506.5 & $\pm$ &     4756.8 &   241585.3 & $\pm$ &     4778.2 \\
S13  &   91.6 &  42.20  &   530163.5 & $\pm$ &     5766.6 &   553280.7 & $\pm$ &     6018.0 &   555762.4 & $\pm$ &     6045.0 \\
S14  & 1465.9 &  5.60   &    70393.5 & $\pm$ &     9402.2 &    73462.9 & $\pm$ &     9812.1 &    73792.4 & $\pm$ &     9856.1 \\
S17  &   66.1 &  67.35  &   846783.7 & $\pm$ &    40846.2 &   883706.7 & $\pm$ &    42627.3 &   887670.5 & $\pm$ &    42818.5 \\
S18  &  107.1 &  34.71  &   436051.9 & $\pm$ &     8874.8 &   455065.4 & $\pm$ &     9261.8 &   457106.6 & $\pm$ &     9303.4 \\
S19  &   87.1 &  72.62  &   914921.4 & $\pm$ &   349248.6 &   954815.5 & $\pm$ &   364477.2 &   959098.2 & $\pm$ &   366112.0 \\
S21  &  217.4 &  19.18  &   240918.1 & $\pm$ &    12090.0 &   251423.0 & $\pm$ &    12617.2 &   252550.7 & $\pm$ &    12673.7 \\
S22  &   19.0 &  456.10 &  5744036.1 & $\pm$ &  2267999.6 &  5994498.0 & $\pm$ &  2366893.0 &  6021386.0 & $\pm$ &  2377509.6 \\
S23  &  114.1 &  34.54  &   435270.2 & $\pm$ &   105542.1 &   454249.7 & $\pm$ &   110144.2 &   456287.2 & $\pm$ &   110638.2 \\
S24  &  107.5 &  97.28  &  1222592.3 & $\pm$ &   134501.7 &  1275901.9 & $\pm$ &   140366.5 &  1281624.9 & $\pm$ &   140996.1 \\
S29  &   98.5 &  57.33  &   720933.8 & $\pm$ &   135104.2 &   752369.2 & $\pm$ &   140995.3 &   755743.9 & $\pm$ &   141627.7 \\
S31  &   63.3 &  82.46  &  1041902.1 & $\pm$ &    37885.2 &  1087333.0 & $\pm$ &    39537.1 &  1092210.2 & $\pm$ &    39714.5 \\
S33  &   47.9 &  135.83 &  1708725.8 & $\pm$ &   259656.2 &  1783232.7 & $\pm$ &   270978.2 &  1791231.3 & $\pm$ &   272193.7 \\
S38  &  427.5 &  8.31   &   104612.8 & $\pm$ &      496.8 &   109174.3 & $\pm$ &      518.4 &   109664.0 & $\pm$ &      520.7 \\
S39  &  364.5 &  19.25  &   242300.3 & $\pm$ &    19531.2 &   252865.5 & $\pm$ &    20382.8 &   253999.7 & $\pm$ &    20474.2 \\
S42  &   30.8 &  250.45 &  3139773.4 & $\pm$ &  1219015.1 &  3276679.5 & $\pm$ &  1272168.8 &  3291376.9 & $\pm$ &  1277875.0 \\
S54  &   81.5 &  144.02 &  1800188.1 & $\pm$ &  2886287.8 &  1878683.2 & $\pm$ &  3012140.9 &  1887109.9 & $\pm$ &  3025651.7 \\
S55  &  382.8 &  7.38   &    92621.1 & $\pm$ &     2894.4 &    96659.7 & $\pm$ &     3020.7 &    97093.3 & $\pm$ &     3034.2 \\
S60  &  105.5 &  50.59  &   635974.4 & $\pm$ &    24431.1 &   663705.3 & $\pm$ &    25496.4 &   666682.4 & $\pm$ &    25610.8 \\
S66  &   13.4 &  655.82 &  8246493.5 & $\pm$ &   851590.8 &  8606072.0 & $\pm$ &   888723.4 &  8644674.1 & $\pm$ &   892709.7 \\
S67  &   19.3 &  402.94 &  5066595.9 & $\pm$ &   314331.6 &  5287518.8 & $\pm$ &   328037.6 &  5311235.7 & $\pm$ &   329509.0 \\
S71  &  106.2 &  100.28 &  1261680.1 & $\pm$ &   193119.5 &  1316694.1 & $\pm$ &   201540.2 &  1322600.1 & $\pm$ &   202444.2 \\
S83  &   15.3 &  589.30 &  7410617.8 & $\pm$ &  1768536.9 &  7733749.0 & $\pm$ &  1845651.9 &  7768438.3 & $\pm$ &  1853930.4 \\
S85  &   11.0 & 1772.21 & 22152053.5 & $\pm$ & 33765230.9 & 23117967.4 & $\pm$ & 35237523.6 & 23221661.8 & $\pm$ & 35395579.5 \\
S87  &    7.6 & 1577.89 & 19792626.5 & $\pm$ &  1922704.9 & 20655660.4 & $\pm$ &  2006542.2 & 20748310.3 & $\pm$ &  2015542.4 \\
S89  &   31.0 & 273.90  &  3439128.4 & $\pm$ &   474176.5 &  3589087.5 & $\pm$ &   494852.4 &  3605186.1 & $\pm$ &   497072.0 \\
S91  &   11.4 & 891.25  & 11199826.5 & $\pm$ &   970523.3 & 11688181.6 & $\pm$ &  1012841.8 & 11740608.3 & $\pm$ &  1017384.8 \\
S96  &   13.6 & 646.91  &  8134554.8 & $\pm$ &   512145.9 &  8489252.4 & $\pm$ &   534477.4 &  8527330.4 & $\pm$ &   536874.7 \\
S97  &    9.7 & 1151.43 & 14531570.5 & $\pm$ &  5278240.7 & 15165202.3 & $\pm$ &  5508392.1 & 15233225.0 & $\pm$ &  5533099.7 \\
S145 &   23.6 & 343.33  &  4332795.7 & $\pm$ &  2127712.2 &  4521722.1 & $\pm$ &  2220488.5 &  4542004.0 & $\pm$ &  2230448.4 \\
S175 & 1812.0 & 6.30    &    79183.3 & $\pm$ &    19172.5 &    82636.0 & $\pm$ &    20008.5 &    83006.7 & $\pm$ &    20098.3 \\
R34  &   18.6 & 589.73  &  7427033.2 & $\pm$ &  2111171.4 &  7750880.2 & $\pm$ &  2203226.5 &  7785646.3 & $\pm$ &  2213108.9 \\
R44  &    5.5 & 2579.33 & 33005557.9 & $\pm$ & 21665177.6 & 34444725.9 & $\pm$ & 22609861.9 & 34599225.9 & $\pm$ & 22711277.1 \\
\hline
\end{tabular}
\normalsize
\label{tab2}
\end{table}

The dependence of the gravity range parameter $\Lambda$ on the observed orbital periods $P$ and eccentricities $e$ of a number of S-stars is shown in Fig. \ref{fig02} for 4 different values of parameter $\delta$: $\delta$ = 0.1, 1, 10 and 100, respectively (see also Table \ref{tab1}). Like in Fig. \ref{fig01}, it is evident that the larger values of $\Lambda$ are obtained for S-stars with larger orbital periods $P$ and lower eccentricities $e$ as it follows from Eq. (\ref{equ3.9}). We can see that the estimate of $\Lambda$ is very sensitive to the value of Yukawa strength parameter $\delta$, and that the larger values of $\Lambda$ are obtained for higher values of $\delta$. For $\delta \gtrsim 10$, there is very small influence on $\Lambda$ compared to case $\delta$ = 1. Also, we can conclude that the gravity parameters $\delta$ and $\Lambda$ are mutually dependent only for smaller values of $\delta$.

\begin{figure}[ht!]
\centering
\includegraphics[width=0.49\textwidth]{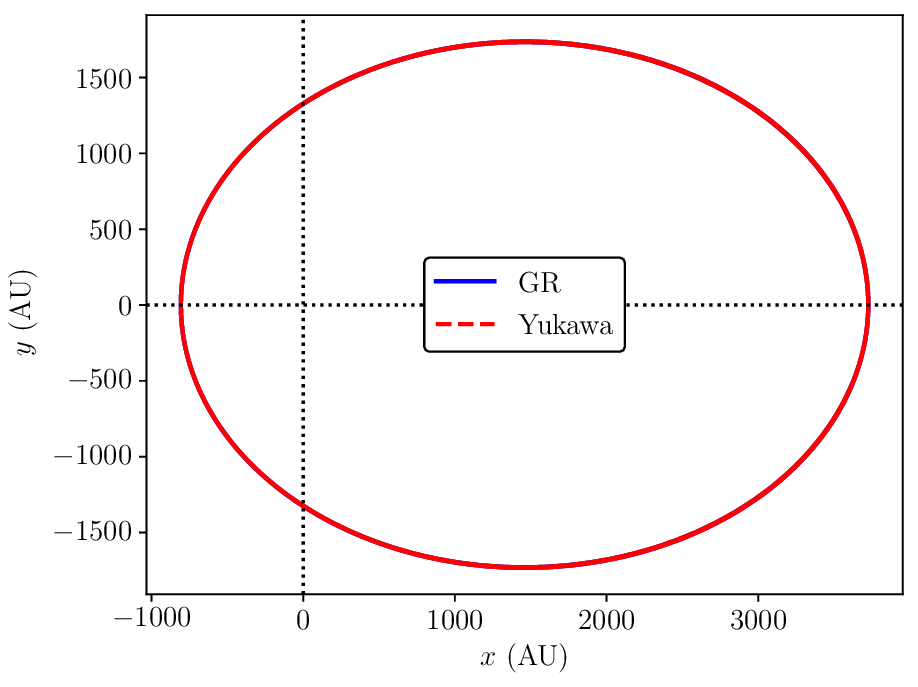}\hfill
\includegraphics[width=0.5\textwidth]{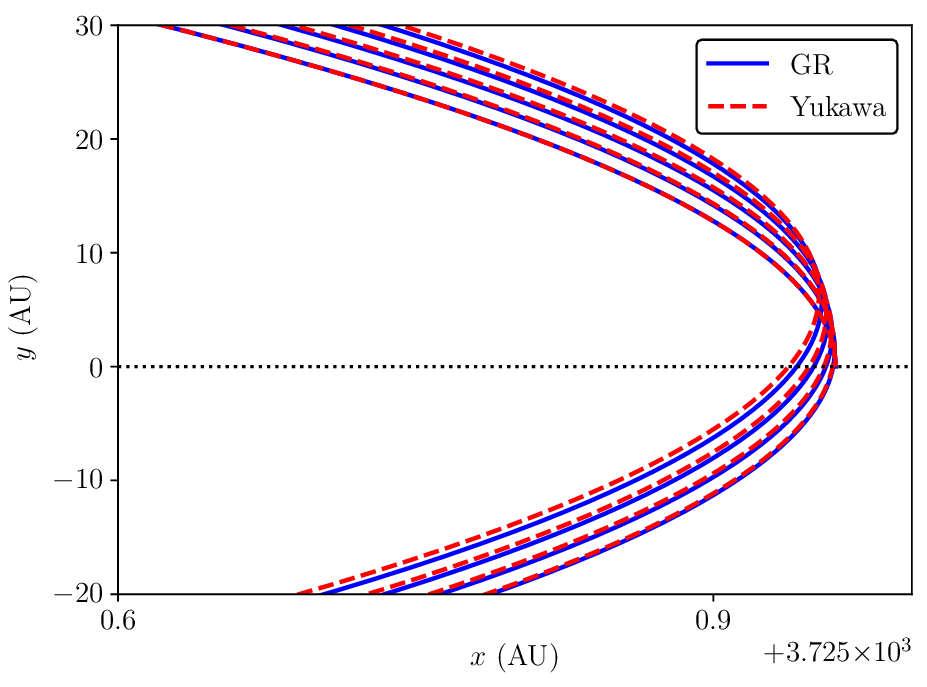}\\
\includegraphics[width=0.5\textwidth]{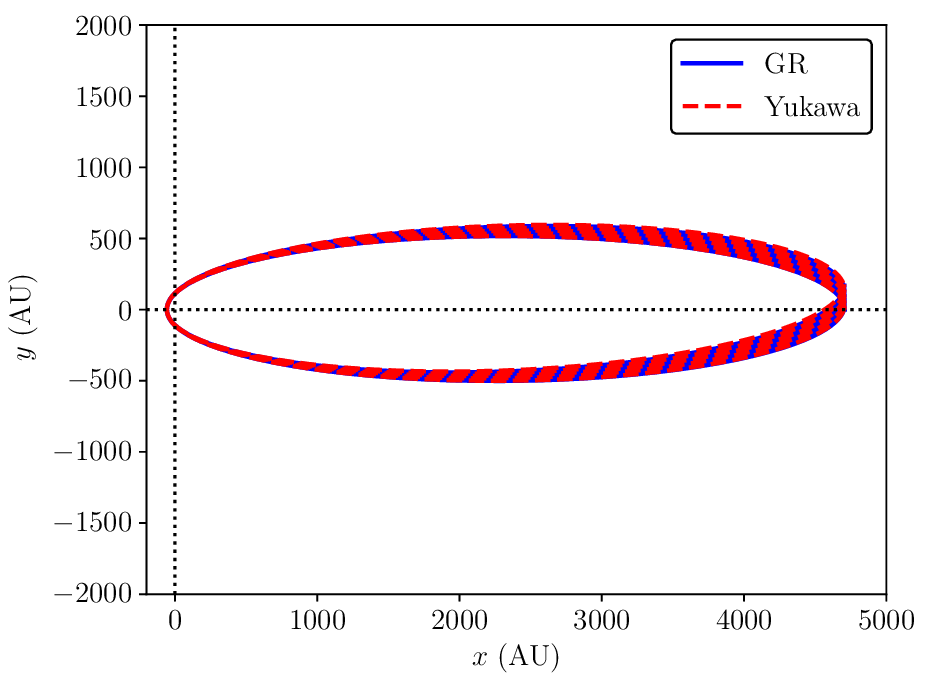}\hfill
\includegraphics[width=0.5\textwidth]{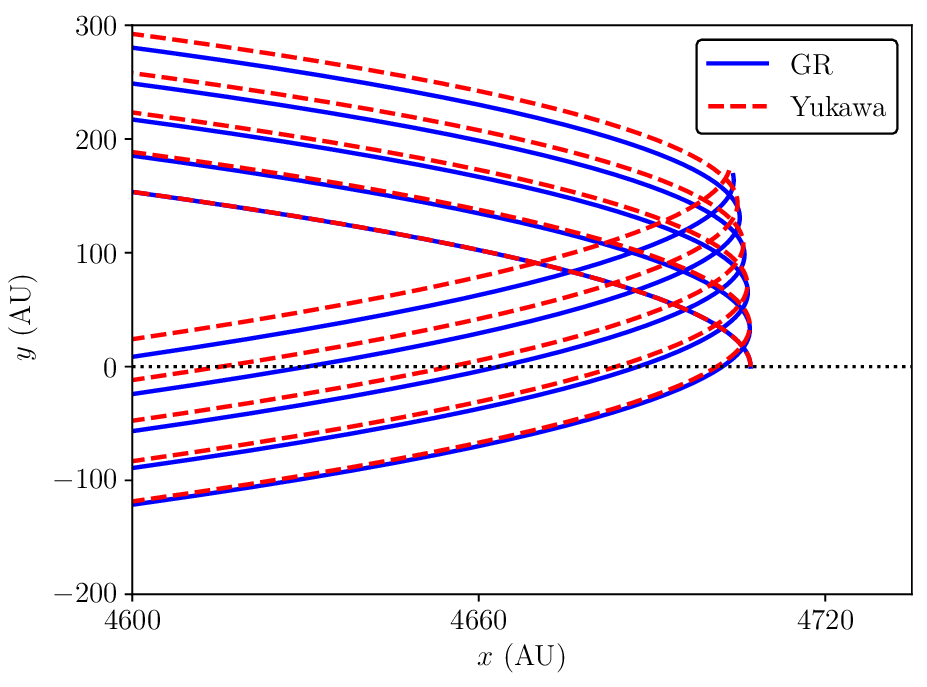}\\
\includegraphics[width=0.46\textwidth]{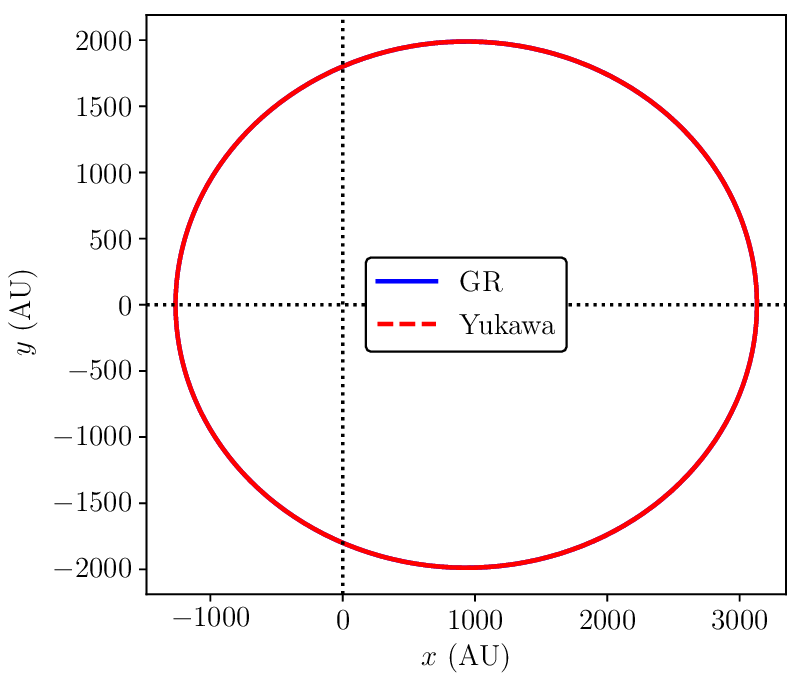}\hfill
\includegraphics[width=0.53\textwidth]{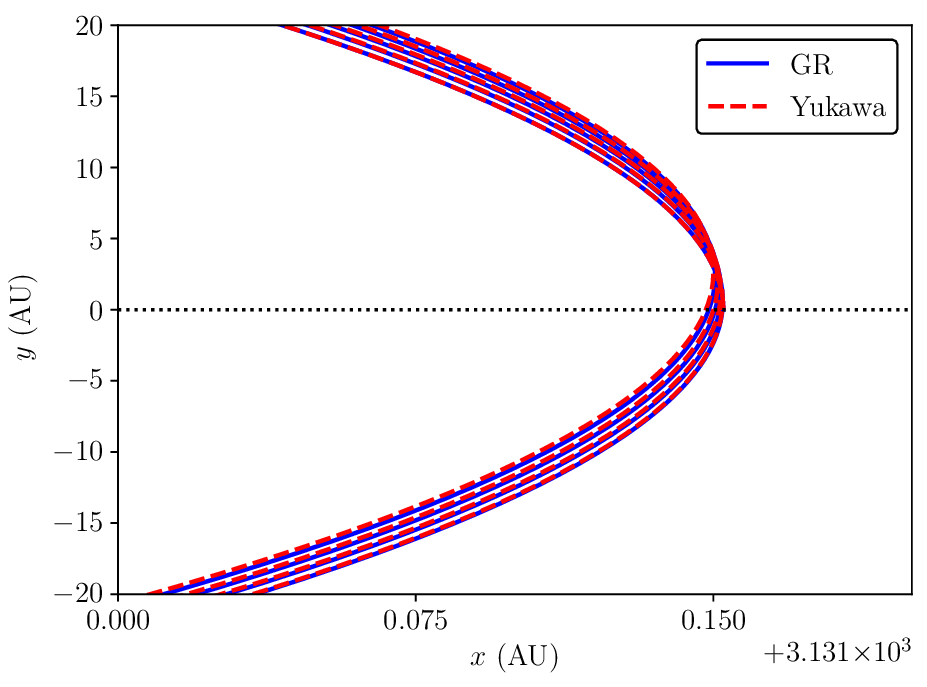}
\caption{Comparison between the simulated orbits of S9 (top), S14 (middle) and S13 star (botom), during five orbital periods in GR (blue solid line) and in Yukawa gravity (red dashed line) for $\delta = 1$. The corresponding values of the range of Yukawa interaction are: $\Lambda$ = 320000.8, 52205.2 and 393179.8 AU, respectively. The left panels show full stellar orbits, while the regions around the apocenter are zoomed in the right panels for better insight.}
\label{fig03}
\end{figure}

Fig. \ref{fig03} shows comparisons between the simulated orbits of S9, S14 and S13 stars in both GR and Yukawa gravity, obtained by numerical integration of the corresponding equations of motion (\ref{equ3.1}) and (\ref{equ3.2}) during their five orbital periods and for $\delta = 1$. The left panels in Fig. \ref{fig03} show the full stellar orbits, while the regions around their apocenters are zoomed in the right panels for better insight. These three stars are chosen due to their drastically different eccentricities, as well as due to their orbital periods around 50 yr, which makes these stars the good candidates for future monitoring with large observational facilities.

Periods of S9, S14 and S13 stars are $P$ = 51.3, 55.3 and 49.0 yr and their eccentricities are $e$ = 0.644, 0.9761, and 0.425, respectively. As it can be seen from Fig. \ref{fig03}, there are a small deviations between GR and Yukawa orbits in the case of these three S-stars. Orbits are obtained for $\Lambda$ = 320000.8, 52205.2 and 393179.8 AU, respectively. These deviations are expected \cite{abut20}. Assuming that the range of Yukawa gravity $\Lambda$ is fixed to its value corresponding to $\delta=1$ (e.g. $\Lambda$ = 52205.2 AU in the case of S14 star), and that instead of $\delta=1$ we use different values (e.g. $\delta$ = 0.1, 10 or 1000), these deviations become larger (see the case for S14 star presented in Fig. 4). The discrepancy between the simulated orbits in Yukawa gravity and in GR becomes noticeable for $\delta$ > 0.1 and rises with increase of $\delta$ and then saturates and tends to become nearly constant for $\delta > 1$. Also, if $\Lambda$ is fixed, and $\delta$ is varied, orbits in Yukawa gravity change significantly in the range $0.1 <\delta < 1$. Outside of this interval for $\delta$, influence of $\delta$ on S-star orbit is very small. Thus, one can conclude that the influence of strength of Yukawa interaction $\delta$ on stellar orbits is noticeable for range $0.1 <\delta < 1$, especially in case of S-stars with high orbital eccentricities $e$.

\begin{figure}[ht!]
\centering
\includegraphics[width=0.5\textwidth]{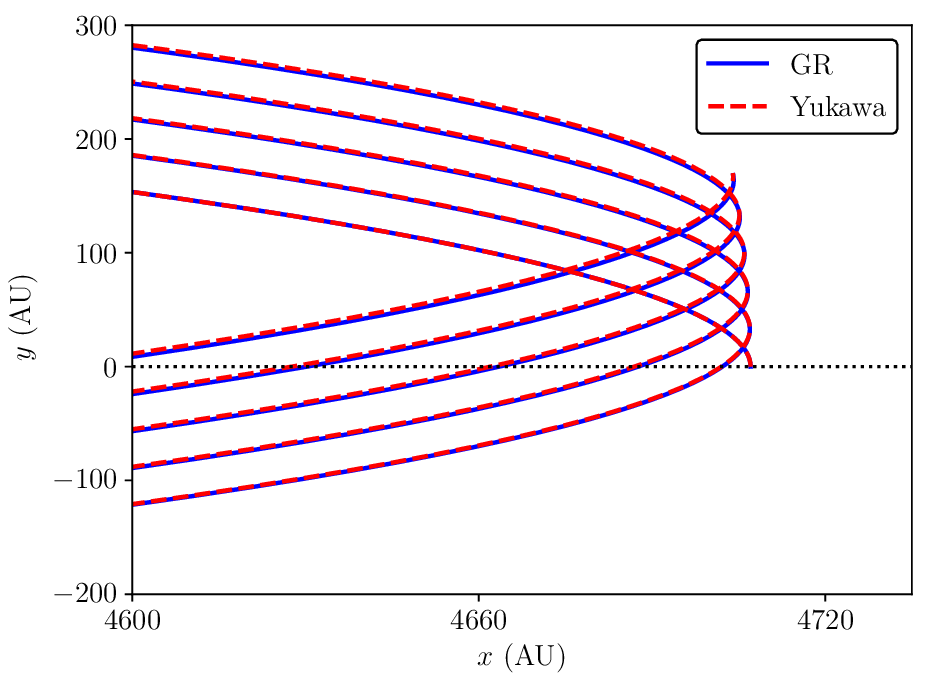}\hfill
\includegraphics[width=0.5\textwidth]{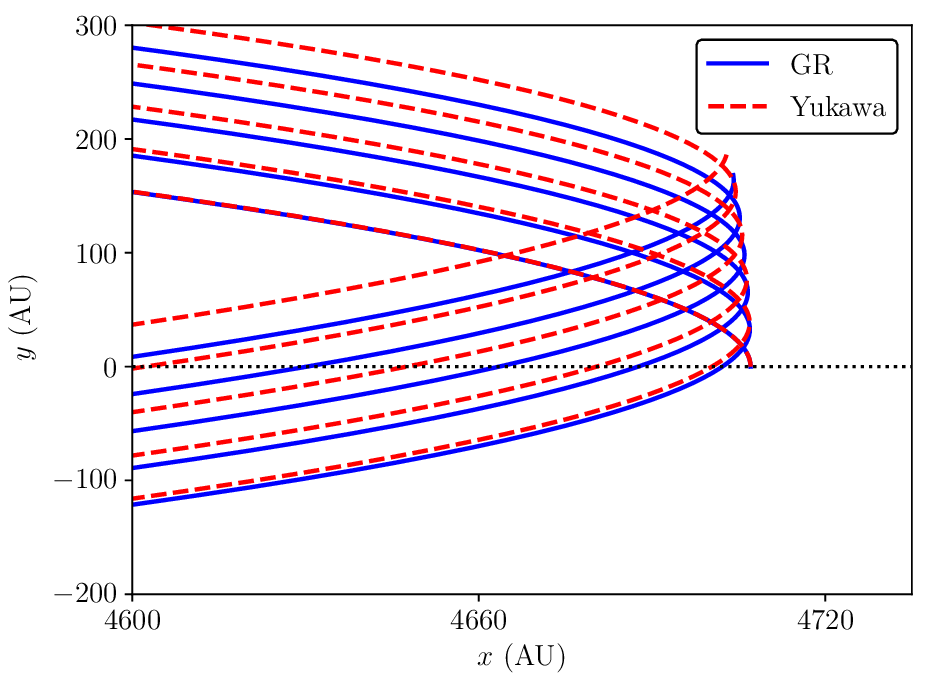}\\
\includegraphics[width=0.5\textwidth]{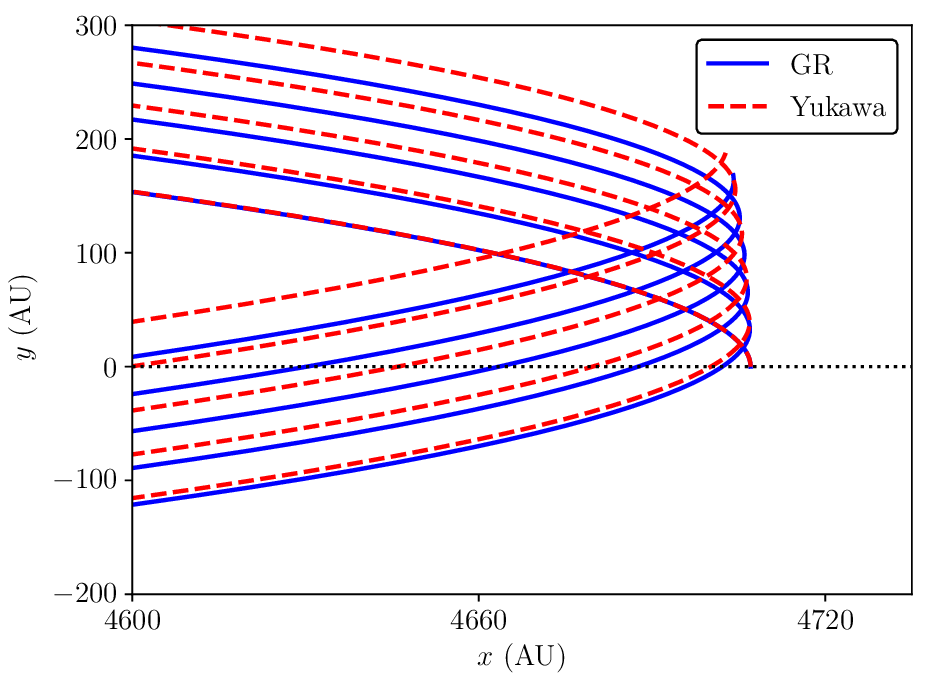}\hfill
\includegraphics[width=0.5\textwidth]{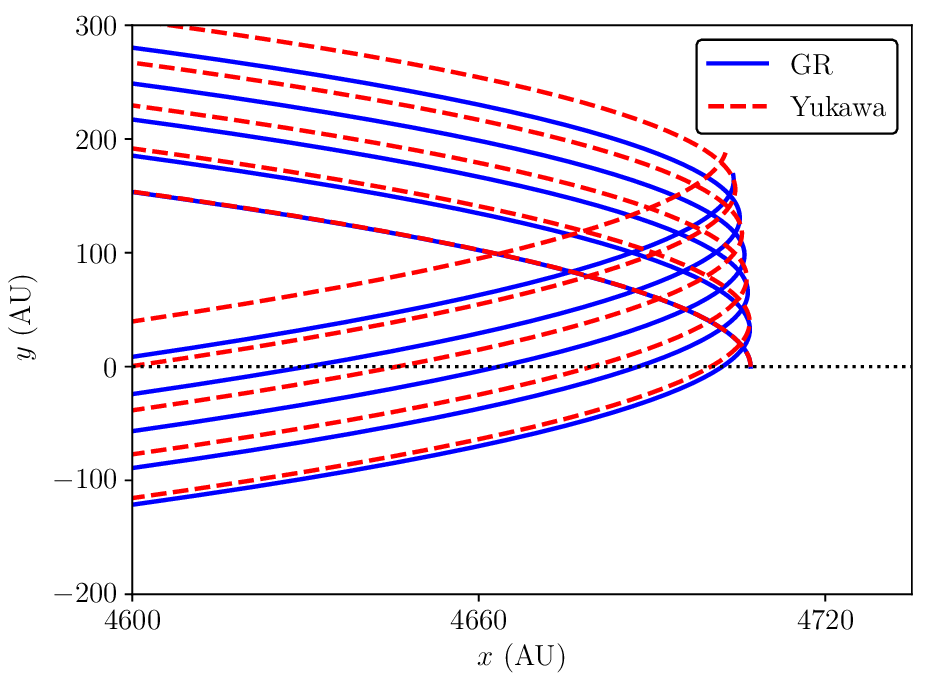}
\caption{Comparisons between the simulated orbits of S14 star in GR (blue solid line) and in Yukawa gravity (red dashed line) for the following 4 different values of $\delta$: $\delta = 0.1$ (top left), $\delta = 10$ (top right), $\delta = 100$ (bottom left) and $\delta = 1000$ (bottom right). The orbits are calculated during five orbital periods, and their zoomed parts around the apocenter are presented for better insight. The corresponding value of $\Lambda$ = 52205.2 AU.}
\label{fig04}
\end{figure}

\subsection{Constrains on Yukawa gravity parameters from Markov chain Monte Carlo simulations}

In this subsection we estimated 68\% confidence region for the Yukawa gravity parameters $\delta$ and $\Lambda$ using Markov chain Monte Carlo (MCMC) \cite{fore13,audr13,shar17,hogg18,bork22b,hogg10}. MCMC methods provide very efficient sampling approximations to the posterior probability density function in parameter spaces \cite{fore13}. We used MIT licensed pure-Python implementation of Goodman \& Weare's Affine Invariant Markov chain Monte Carlo Ensemble sampler (\url{https://emcee.readthedocs.io/en/stable/}). The explanation of the \texttt{emcee} algorithm and its implementation in detail are given in \cite{fore13}. We obtained the maximum likelihood values of Yukawa gravity parameters using the \texttt{optimize.minimize} module from SciPy for maximization of their likelihood function. After that we used these maximum likelihood values of the parameters as a starting point for our MCMC simulations, which we performed in order to estimate the posterior probability distributions for the Yukawa gravity parameters. To perform these simulations, we used $0 < \delta < 10^3$ and $10^3$AU $ < \Lambda < 10^6$AU as our priors.

\begin{figure}[ht!]
\centering
\fbox{\includegraphics[width=0.48\textwidth]{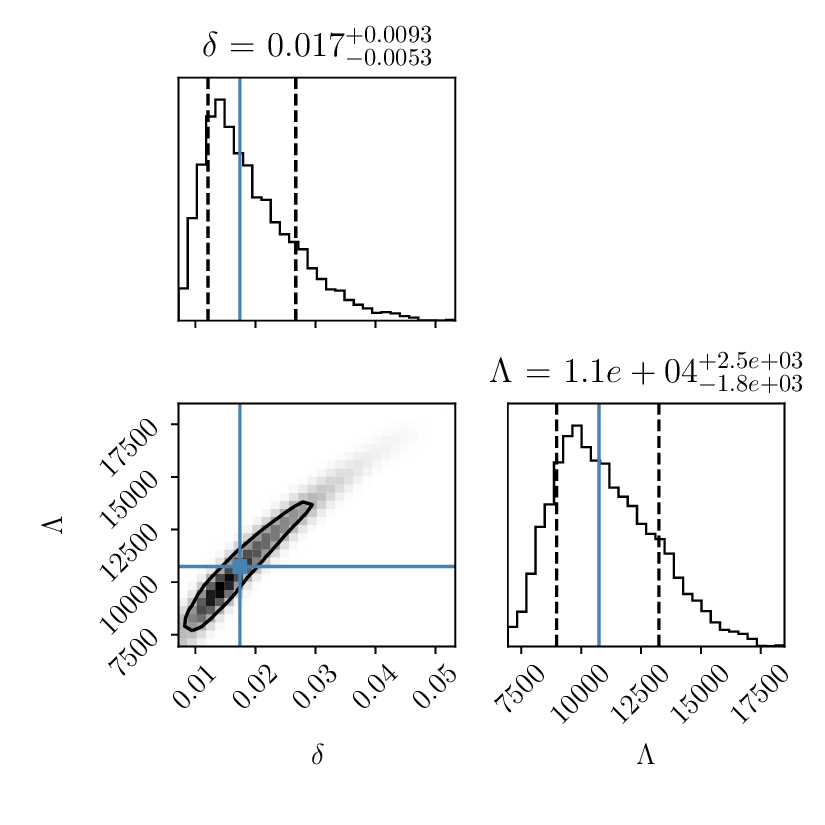}}\hfill
\fbox{\includegraphics[width=0.48\textwidth]{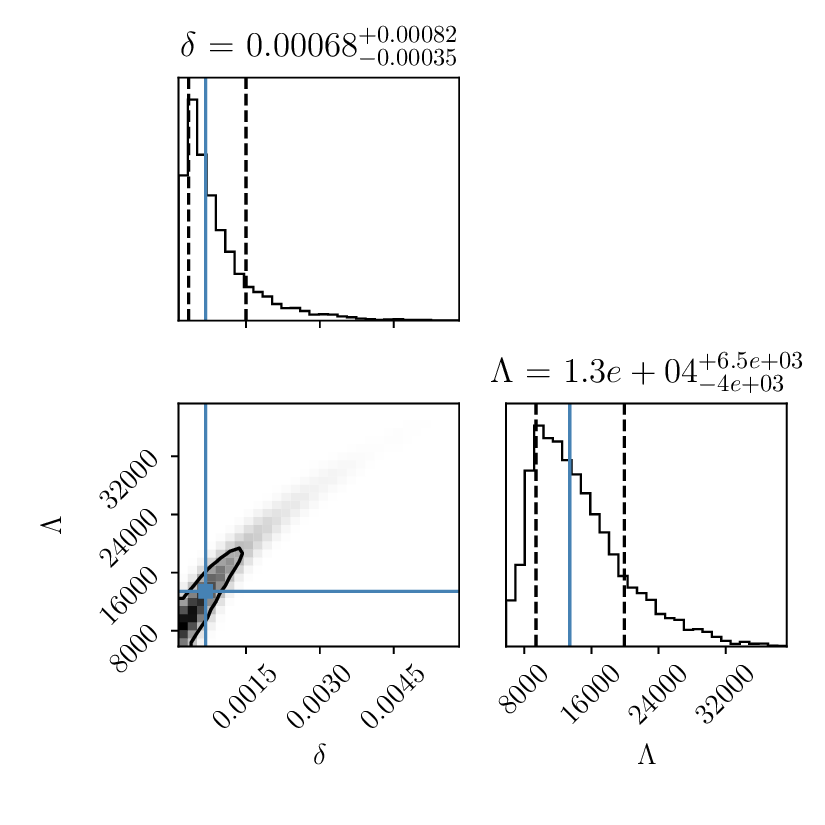}}\vspace*{0.1cm}
\fbox{\includegraphics[width=0.48\textwidth]{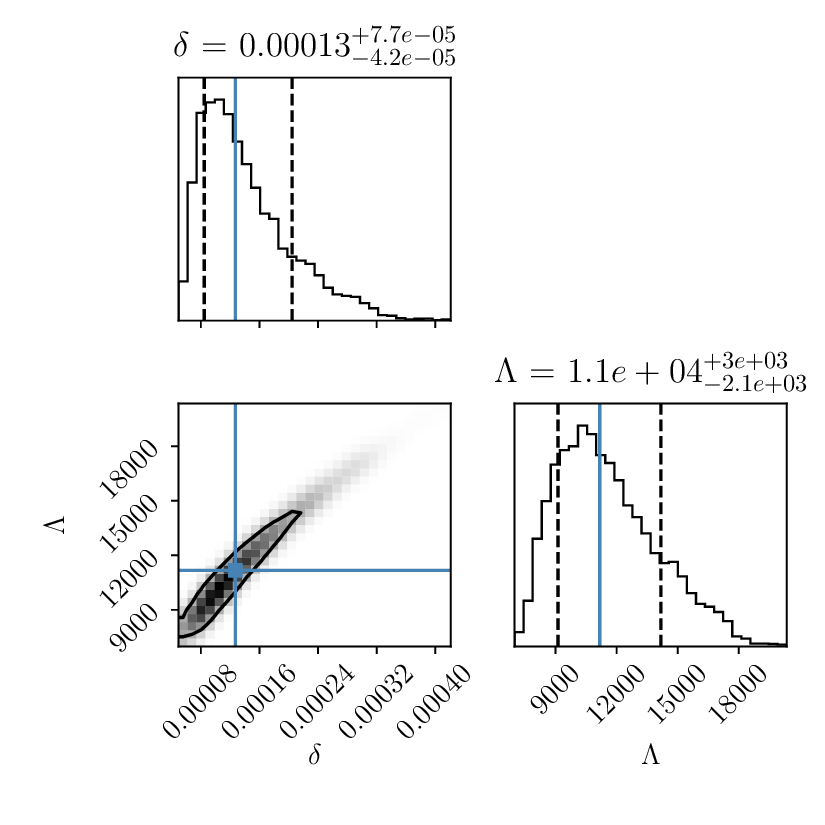}}\hfill
\fbox{\includegraphics[width=0.48\textwidth]{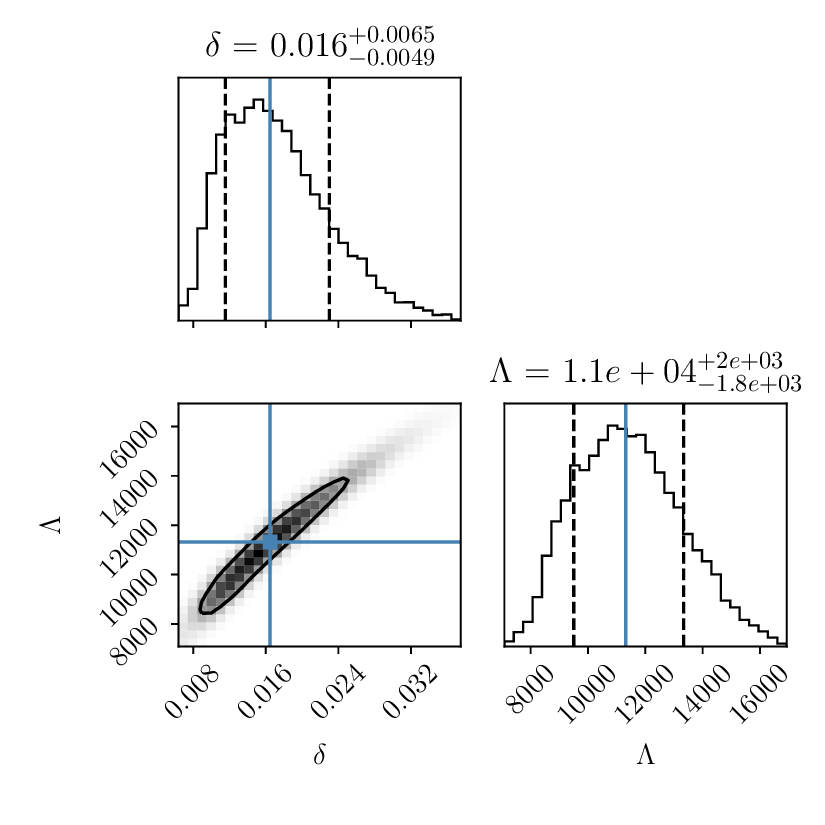}}
\caption{The posterior probability distributions and 68\% confidence levels (closed contours) for the Yukawa gravity parameters in the case of S-stars, obtained by MCMC simulations. Top left panel corresponds to the following S-stars: S2, S12, S14, S21, S38, S39, S55 and S175, for which 5 yr $< P^{*} <$ 25 yr (see Table \ref{tab1}), top right panel corresponds to S8, S9, S13, S18 and S23 star for which 25 yr $ < P^{*} <$ 50 yr, bottom left panel corresponds to S4, S6, S17, S19, S24, S29, S31 and S60 star for which 50 yr $ < P^{*} <$ 100 yr, while bottom right panel correspond to all S-stars from Table \ref{tab1}. The designated best-fit values of Yukawa gravity parameters and their uncertainties were obtained from the 16th, 50th and 84th percentiles of the corresponding posterior probability distributions (vertical lines in the histograms).}
\label{fig05}
\end{figure}

In order to test the robustness of our MCMC analysis, as well as the possible influence of the scale of each gravitational system on the obtained estimates, we performed the MCMC simulations not only for all S-stars from Table \ref{tab1}, but also for their following three subsamples, defined by scale criterion $P^{*}$ (see Table \ref{tab1} for its particular values):
\begin{enumerate}
\setlength{\itemsep}{0em}
\item subsample where 5 yr $ < P^{*} <$ 25 yr, containing S2, S12, S14, S21, S38, S39, S55 and S175 star,
\item subsample where 25 yr $ < P^{*} <$ 50 yr, containing S8, S9, S13, S18 and S23 star,
\item subsample where 50 yr $ < P^{*} <$ 100 yr, containing S4, S6, S17, S19, S24, S29, S31 and S60 star.
\end{enumerate}

Fig. \ref{fig05} represents the obtained posterior probability distributions of the parameters of Yukawa gravity model ($\delta$ and $\Lambda$), where the contours represent their 68\% confidence levels. In the case of the first subsample, the following best-fit values of Yukawa gravity parameters and their uncertainties were obtained from the 16th, 50th and 84th percentiles of the posterior probability distributions: $\delta = 0.017^{+0.0093}_{-0.0053}$ and $\Lambda = 11000^{+2500}_{-1800}$ AU. (see top left panel of Fig. \ref{fig05}). The following results were obtained in the case of the second subsample: $\delta = 0.00068^{+0.00082}_{-0.00035}$ and $\Lambda = 13000^{+6500}_{-4000}$ AU (see top right panel of Fig. \ref{fig05}), while in the case of the third subsample we got the following values: $\delta = 0.00013^{+0.000077}_{-0.000042}$ and $\Lambda = 11000^{+3000}_{-2100}$ AU (see bottom left panel of Fig. \ref{fig05}). Bottom right panel of Fig. \ref{fig05} shows the corresponding results obtained for for all S-stars from Table \ref{tab1}, in which case the following best-fit values were obtained: $\delta = 0.016^{+0.0065}_{-0.0049}$ and $\Lambda = 11000^{+2000}_{-1800}$ AU.

It can be seen from these results that the scale of gravitational systems also plays a significant role, since the smaller values of universal constant $\delta$ are obtained for the subsamples of S-stars with higher values of scale criterion $P^{*}$. Therefore, the S-stars with large $P^{*}$ cause that the value of $\delta$ for whole sample (bottom right panel of Fig. \ref{fig05}) is slightly less than its corresponding value for the first subsample with smallest $P^{*}$ (top left panel of Fig. \ref{fig05}). All these results obtained by MCMC simulations are consistent with our previous analyses concerning the constraints on the Yukawa gravity parameters
\cite{bork13,zakh16a,zakh18a}.

Also, constraints of Yukawa gravity parameters are presented in recent following papers: \cite{rham17,will18,dadd21,mart21,moni22,beni22b,dong22}. In the references \cite{mart21,moni22} authors use publicly available data for the S2 star, integrated geodesics and use MCMC analysis method. They discussed the usage of S2-stars observations to constrain a Yukawa-like gravitational potential and to constrain parameter space of MOdified gravity (MOG). Authors analyzed observational data and concluded that the orbital precessions of the S2 star is in good agreement with the corresponding prediction of GR. In paper \cite{dadd21} author performs data analysis in the framework of Yukawa gravity model by solving geodesic equation, and concluded that the orbital precessions of the S2, S38 and S55 stars are close to the corresponding prediction of GR for these stars, i.e. no significant departure from GR is detected. Results of above mentioned papers for Yukawa gravity parameters $\delta$ and $\Lambda$ are in good agreement with our findings from this paper (obtained using MCMC analysis of crytical period P$^{*}$ for all S-stars).

In paper \cite{rham17}, the different graviton mass bounds obtained from massive potentials, Yukawa like and non-Yukawa like, are reviewed. The graviton mass bound from Solar System, Clusters and weak lensing with bound from gravitational waves GW150914 are compared. In Ref. \cite{will18} the solar-system data in case of Yukawa form of gravitation potential are analyzed and obtained bounds of graviton mass. There are also constrained Yukawa gravitational parameters using data from paper \cite{pitj13}. In paper \cite{beni22b} the scalar-tensor theories in the strong gravity regime around BH at the centre of our galaxy are investigated, S2 star fit is presented as well as the fit of the combined data of S2 and S38. The analysis of Yukawa gravity parameters with pulsars around Sgr A$^{*}$ is given in \cite{dong22}.

We compare our results with another different constraints of the Yukawa parameters for some S-stars and from Solar System results. These two systems have approximately comparable scale sizes, i.e. scale of the S2 orbit is of the same order as the Solar System. In paper \cite{will18} the author presented the latest analyses of Solar System data, and showed that the best bound on $\Lambda$ comes from the perihelion advance of Mars, giving thus the range $\Lambda > (1.2 - 2.2) \times 10^{14}$ km depending on the specific analysis. With the calculation performed using data from Table 4 of \cite{pitj13}, in \cite{will18} the results for range of $\Lambda$ (all given in $10^{14}$ km) are the following: Mercury 0.18; Venus 0.28; Earth 0.88; Mars 2.21; Jupiter 0.11 and Saturn 0.98. In reference \cite{beni22} best-fit values were obtained for the parameter values of the Yukawa potential from the Solar System precession: $\delta = 3.863^{+0.373}_{-0.373}$ $10^{-3}$ and $1/\Lambda= 1.066^{+0.6074}_{-0.6074}$ $10^{-3}$ AU$^{-1}$. In paper \cite{dadd21} best-fit values were obtained for the S2 star data: $\delta = 0.00^{+1.53}_{-0.36}$, $\Lambda \geq 7059.29$ AU; for S38 star data: $\delta = 0.00^{+1.54}_{-0.34}$, $\Lambda \geq 5731.30$ AU for S55 star data: $\delta = -0.10^{+1.62}_{-0.33}$, $\Lambda  \geq 3511.95$ AU and for multi-star data: $\delta = 0.00^{+1.69}_{-0.52}$, $\Lambda \geq 6336.23$ AU. In reference \cite{mart21} best-fit values were obtained for the S2 star data, by excluding and including the measurement of the orbital precession of S2 star, respectively: $\delta \geq -0.07$, $\Lambda  \geq 9540$ AU, and $\delta \geq -0.01^{+0.61}_{-0.14}$, $\Lambda \geq 6300$ AU.

All these authors obtain that the range of Yukawa interaction is of the similar order of magnitude, as the results presented here.

\section{Conclusions}

We investigated the influence of strength of Yukawa interaction $\delta$ on the observed stellar orbits of the S-stars. All of these results were obtained assuming that the precession angles of S-stars in Yukawa gravity will have a small deviation from GR prediction (as indicated by GRAVITY Collaboration \cite{abut20}). We believe that Yukawa can provide a fit for the trajectories of S-stars, however assuming that in the future observations Schwarzschild precessions will be close to their GR estimates, we showed that these precessions can be fitted by a set of Yukawa gravity parameters. Our results showed the following:

\begin{itemize}
\setlength{\itemsep}{0em}
\item Strength of Yukawa interaction $\delta$ strongly affects its range $\Lambda$ and vice versa, i.e. these two gravity parameters are mutually correlated for smaller values of $\delta$;
\item Orbits with larger periods $P$, smaller eccentricities $e$ and higher value of strength of Yukawa interaction $\delta$ provide larger values of range of Yukawa interaction $\Lambda$;
\item Strength of Yukawa interaction $\delta$ has noticeable influence on stellar orbits, especially in case of S-stars with high orbital eccentricities $e$. Discrepancy between Yukawa and GR simulated orbits is better visible in case of orbits with higher eccentricity $e$, because value of precession is then higher and it is more easy to detect discrepancy from GR orbit. However, values of $\Lambda$ are larger for lower eccentricities $e$;
\item In the case of the fixed range of Yukawa interaction $\Lambda$, discrepancy between the simulated orbits in Yukawa gravity and in GR becomes noticeable for $\delta > 0.1$ and rises with increase of $\delta$ and then saturates and tends to become nearly constant for $\delta > 1$;
\item Our MCMC simulations resulted with the following best-fit values and uncertainties of Yukawa gravity parameters in the case of the sample of S-stars from Tables \ref{tab1} and \ref{tab2}: $\delta = 0.017^{+0.0093}_{-0.0053}$ and $\Lambda = 11000^{+2500}_{-1800}$ AU.
\item These MCMC simulations also demonstrated that the scale of gravitational systems plays a significant role and that it can have significant influence on the values of both Yukawa gravity parameters. S-stars which represent larger gravitational systems are better described with the lower strength of Yukawa interaction $\delta$, while the opposite holds for the S-stars with orbits of smaller scales.
\item We defined scale criterion $P^{*}$, according to which we divided samples of all studied S-stars into three subsamples. These results showed that the scale of the gravitational system (described by $P^{*}$) is very sensitive to the values of strength of Yukawa interaction $\delta$. Therefore, the quantity $P^{*}$ can be used as a criterion for classification of the gravitational systems according to their scales, in the frame of Yukawa gravity.
\end{itemize}

It is very important to investigate the influence of strength on Yukawa interaction on observed stellar orbits around GC in the frame of the Yukawa-type gravity theories because it represents an excellent way for probing and testing the predictions of gravity theories. Also, we believe that our current estimates for Yukawa gravity parameters can be significantly improved with new observations of trajectories of bright stars near the Galactic Center like GRAVITY \cite{blin15}, E-ELT \cite{eelt10} and TMT \cite{tmt15}.

\acknowledgments

This work is supported by Ministry of Science, Technological Development and Innovations of the Republic of Serbia through the Project contracts No. 451-03-47/2023-01/200002 and 451-03-47/2023-01/200017.


\end{document}